\documentclass[fleqn,usenatbib]{mnras}

\usepackage{newtxtext,newtxmath}

\usepackage[T1]{fontenc}
\usepackage{ae,aecompl}

\usepackage{graphicx}	
\usepackage{amsmath}	
\usepackage{amssymb}	
\usepackage{subfig}     
\usepackage{soul}
\usepackage{float}

\newcommand{\ra}[1]{\renewcommand{\arraystretch}{#1}}

\newcommand{\Vobs}{V_{\mathrm{obs}}}
\newcommand{\Vflat}{V_{\mathrm{flat}}}

\newcommand{\Vbary}{V_{\mathrm{bar}}}
\newcommand{\Vb}{V_{\mathrm{bar}}}

\newcommand{\Rbary}{R_{\mathrm{bar}}}

\newcommand{\SBbar}{\Sigma_{\mathrm{bar}}}

\newcommand{\Rd}{R_{\mathrm{d}}}
\newcommand{\Vdisc}{V_{\mathrm{disc}}}

\newcommand{\MtLdisc}{\Upsilon_{\mathrm{disc}}}
\newcommand{\maxMtLdisc}{\Gamma_{\mathrm{disc}}}

\newcommand{\MtLbul}{\Upsilon_{\mathrm{bul}}}
\newcommand{\maxMtLbul}{\Gamma_{\mathrm{bul}}}

\newcommand{\Vgas}{V_{\mathrm{gas}}}

\newcommand{\VbVp}{V_{\rm bar}/V_{\rm obs}}

\newcommand{\spacepm}[1]{\hspace{#1 pt}\pm \hspace{#1 pt}}

\usepackage[dvipsnames]{xcolor}

\title[A New Algorithm to Quantify Maximum Discs in Galaxies]{A New Algorithm to Quantify Maximum Discs in Galaxies}

\author[Starkman et al.]{
Nathaniel Starkman,$^{1}$\thanks{email: nms85@case.edu}
Federico Lelli$^{2}$\thanks{ESO Fellow}
Stacy McGaugh,$^{1}$
and James Schombert$^{3}$
\\
$^{1}$Departments of Physics and Astronomy, Case Western Reserve University, Euclid Ave, Cleveland 44106, USA\\
$^{2}$European Southern Observatory, Karl-Schwarzschield-Strasse 2, Garching bei M\"unchen, Germany\\
$^{3}$Department of Physics, University of Oregon, Willamette Hall, Eugene 97403, USA
}

\date{July 24, 2018}

\pubyear{2018}

\begin{document}
\label{firstpage}
\pagerange{\pageref{firstpage}--\pageref{lastpage}}
\maketitle

\begin{abstract}
Maximum disc decompositions of rotation curves place a dynamical upper limit to the mass attributable to stars in galaxies. The precise definition of this term, however, can be vague and varies in usage. We develop an algorithm to robustly quantify maximum-disc mass models and apply it to 153 galaxies from the SPARC database. Our automatic procedure recovers classic results from manual decompositions. High-mass, high-surface-brightness galaxies have mean maximum-disc mass-to-light ratios of $\sim 0.7 \;{\mathrm{M}_\odot}/{\mathrm{L}_\odot}$ in the Spitzer 3.6 $\mu$m band, which are close to the expectations from stellar population models, suggesting that these galaxies are nearly maximal. Low-mass, low-surface-brightness galaxies have very high maximum-disc mass-to-light ratios (up to 10 $\mathrm{M}_\odot/\mathrm{L}_\odot$), which are unphysical for standard stellar population models, confirming they are sub-maximal. The maximum-disc mass-to-light ratios are more closely correlated with surface brightness than luminosity. The mean ratio between baryonic and observed velocity at the peak of the baryonic contribution is $\mathrm{V}_{\mathrm{bar}}/{\mathrm{V}_\mathrm{p}} \approx 0.88$, but correlates with surface brightness, so it is unwise to use this mean value to define the maximum disc concept. Our algorithm requires no manual intervention and could be applied to large galaxy samples from future HI surveys with Apertif, Askap, and SKA.
\end{abstract}

\begin{keywords}
galaxies: discs -- galaxies: bulges -- galaxies: structure -- galaxies: luminosity function, mass function -- dark matter
\end{keywords}



\section{Introduction}
\label{sec:introduction}

The concept of maximum disc was important to establishing the presence and severity of mass discrepancies in spiral galaxies \citep{1985ApJ...295..305V}. Several lines of evidence suggest that the discs of the most luminous spirals are nearly maximal \citep{1999ASPC..182..351S,2010AIPC.1240..309F}. These include (i) the correspondence between the inner shapes of the predicted and observed rotation curves \citep{1986RSPTA.320..447V,2000AJ....120.2884P}, (ii) the disc self-gravity needed to drive spiral structure \citep{1987A&A...179...23A, 2003ApJ...586..143K} and to explain gas flows in barred galaxies \citep{2001ApJ...546..931W,2004A&A...424..799P}, and (iii) dynamical friction in barred galaxies: if dark matter were the dominat component at small radii, the bar pattern speed would be unrealistically slow due to frictional drag from the dark matter halo \citep{2000ApJ...543..704D}.

Other lines of evidence, however, points against maximum discs in bright spirals. The DiskMass survey measured the vertical velocity dispersion of disc stars in 30 spiral galaxies and quantified their disc maximality using the dynamical relation between vertical velocity dispersion, disc scale height, and stellar mass-to-light ratio \citep{2011ApJ...739L..47B, Martinsson:2013kj}. They concluded that spiral galaxies are highly submaximal. However, \cite{2016MNRAS.456.1484A} pointed out that the DiskMass survey used disc scale heights of old stellar populations, whereas the vertical velocity dispersions are light-weighted towards K-giants of relatively young ages. \cite{2016MNRAS.456.1484A} concluded that the galaxies from the DiskMass survey may be maximal when these effects are taken into account.

In any case, the maximum disc places a hard upper limit on the stellar mass of a galaxy. This dynamical limit constrains stellar population synthesis models and the galaxy-wide initial mass function (IMF). For example, the maximum disc values of high-surface-brightness (HSB) galaxies are often overshot by models with a Salpeter IMF \citep{2001ApJ...550..212B, 2003ApJS..149..289B} but they are consistent with models using \citet{2003PASP..115..763C} or \citet{2001MNRAS.322..231K} IMFs.

In low-luminosity and low-surface-brightness (LSB) galaxies, the case for maximum disc is less compelling. The dynamically allowed mass-to-light ratio becomes larger than expected for normal stellar populations \citep{1997MNRAS.290..533D}: dimmer galaxies appear to be progressively more dark matter dominated \citep{2005ApJ...632..859M,2011ApJ...729..118S,2016ApJ...816...42M,2016AJ....152..157L,2016ApJ...827L..19L}, at least for mass-to-light ratios that are normal from the perspective of stellar populations \citep{2001ApJ...550..212B,2014PASA...31...36S,2014AJ....148...77M}. It is also conceivable that the IMF becomes systematically bottom-heavy with lower surface brightness. Though low in amplitude, spiral structure is sometimes present in LSB galaxies and implies rather heavy discs \citep{2002dmap.conf...28F,2013AN....334..785S}, as anticipated by \citet{1998ApJ...499...66M}.

The situation at present is mixed, with data supporting both the maximal and sub-maximal arguments. The discs of bright spirals are maximal \citep{1999ASPC..182..351S,2010AIPC.1240..309F}, unless they are not \citep{1999ApJ...513..561C,2011ApJ...739L..47B}. The discs of low surface brightness galaxies are sub-maximal \citep{1997MNRAS.290..533D}, unless they are maximal \citep{2002dmap.conf...28F,2013AN....334..785S}.

A crucial issue is that the precise meaning of maximum disc is not well defined. \citet{2000AJ....120.2884P} fit H$\alpha$ rotation curves leaving no room for dark matter at small radii. This is certainly maximal, but tends to over-shoot the innermost velocity points since the stellar mass distribution cannot explain all of the outermost rotation curve, resulting into an inevitable trade-off between inner and outer radii. Presumably dark matter halos are not hollow at their centres, but once we leave room for some dark matter, the amount required depends on the assumed halo profile: a cuspy halo profile like the NFW \citep{1996ApJ...462..563N} will imply less stellar mass than a cored halo profile \citep[e.g.,][]{1985ApJ...295..305V}. \citet{1997ApJ...483..103S} attempted to quantify maximum disc based largely on the precedent of prior usage, but this seems rather arbitrary \citep{2015ApJ...801L..20C}.

We define an algorithm that matches the inner shape of the rotation curve while not exceeding it. This result places an upper limit on the dynamical mass of stars that can be uniformly applied to large samples, as we illustrate with the Spitzer Photometry and Accurate Rotation Curve (SPARC) database \citep{2016AJ....152..157L}. The code perform these fits with no manual intervention and can be very useful in light of future large HI surveys with Apertif, Askap, and ultimately SKA. The code is made publicly available on the SPARC database (astroweb.cwru.edu/SPARC).

\section{Methods} 
\label{sec:Methods}

A galaxy can be broadly decomposed into a few baryonic components: gas disc, stellar disc, and bulge (if present). The conversion from 21 cm flux to atomic gas mass is specified by the physics of the spin-flip transition, fixing the gas contribution. The conversion from stellar light to stellar mass is less well determined. For bulge and stellar disc, we need to introduce mass-to-light ratios, denoted as $\MtLbul$ and $\MtLdisc$, respectively. The values of $\MtLbul$ and $\MtLdisc$ can be fixed using stellar population synthesis models \citep{2001ApJ...550..212B, 2014PASA...31...36S, 2014AJ....148...77M, 2016AJ....152..157L, 2016ApJ...827L..19L}, but in this work we will use them as free parameters to determine their maximum allowed values from a dynamical perspective.

We develop a fitting scheme that matches the baryonic rotation curve to the observed one at small radii, specifically near the ``turning radius'' where the rotation curves start to approach a flat part. This automatic algorithm does what astronomers have been traditionally doing by eye in estimating maximum disc fits. The algorithm produces good results in disc dominated cases. However, for galaxies with bulges, the mass-to-light ratios of bulge and disc can be degenerated. In these cases, we break the degeneracy by imposing $\MtLbul > \MtLdisc$. This choice is motivated by stellar population synthesis models. To confirm the algorithm's efficacy and estimate the uncertainties, a Markov Chain Monte Carlo (MCMC) routine was also run on all galaxies, giving consistent results.

\subsection{Residual Function} 
\label{sub:residual_function}
We minimize the function $\mathcal{F(R)}$ which modifies standard residuals using a weighting function $w(R)$:
\begin{eqnarray}
	\mathcal{F}(R) = & \dfrac{\Vobs(R) - \Vbary(R)}{\delta\Vobs(R)} \cdot w(R).
\label{eq:Res}
\end{eqnarray}
Here $R$ is the galactocentric radius, $\Vobs$ is the observed rotation velocity, $\delta\Vobs$ is the corresponding measurement error, and $\Vbary$ is the predicted baryonic velocity given by
\begin{eqnarray}
\Vbary^2(R) = \MtLdisc V_{\rm disc}^2 + \MtLbul V_{\rm bulge}^2 + V_{\rm gas}^2.
\end{eqnarray}
$V_{\rm gas}$, $V_{\rm disc}$, and $V_{\rm bul}$ are the velocity contributions from gas disc, stellar disc, and bulge, respectively. The free parameters are $\MtLdisc$ and $\MtLbul$. In SPARC only 32 galaxies out of 175 have significant bulges \citep[see][for details]{2016AJ....152..157L}.

The weighting function $w(R)$ is meant to over-weight the inner rising part of the rotation curve and under-weight the outer flat part, where the declining contribution from baryons cannot match the observed rotation curve. To achieve this goal, we use a modified log-normal function. A standard log-normal function peaks early and quickly tails off, emphasizing the inner radii before the rotation curve starts to flatten out. However, it does not emphasize the very innermost radii, for which there are a balance of factors to consider. On the one hand, the innermost radii should contribute most to a maximal fit, before $V_{\rm bar}$ starts to decline and cannot reproduce $V_{\rm obs}$ any more. This is especially true for galaxies with bulges: under-weighting the inner region neglects the bulge. On the other hand, the very innermost points are sometimes uncertain due to several reasons: (i) beam smearing effects and non-circular motions may affect the inner values of $\Vobs$, (ii) non-spherical bulges and non-axisymmetric structures like bars may affect the inner values of $\Vbary$, and (iii) pressure support may be important in the inner parts of low-mass galaxies but the resulting corrections to $\Vobs$ are often uncertain. To balance the observational errors with the need of a maximal fit, we modify the standard log-normal function to achieve the weighting function $w(R)$, which is constant until the maximum of the log normal distribution and then tails off like a log normal.

A log normal probability density function is characterized by a mean and variance. To retain sensitivity to the sample distribution while weighting the region of interest, both the mean and variance of the log-normal are set to the log of the mean and variance of the sampled radii in kpc.

\begin{figure*}
	\includegraphics[width=1.0\textwidth]{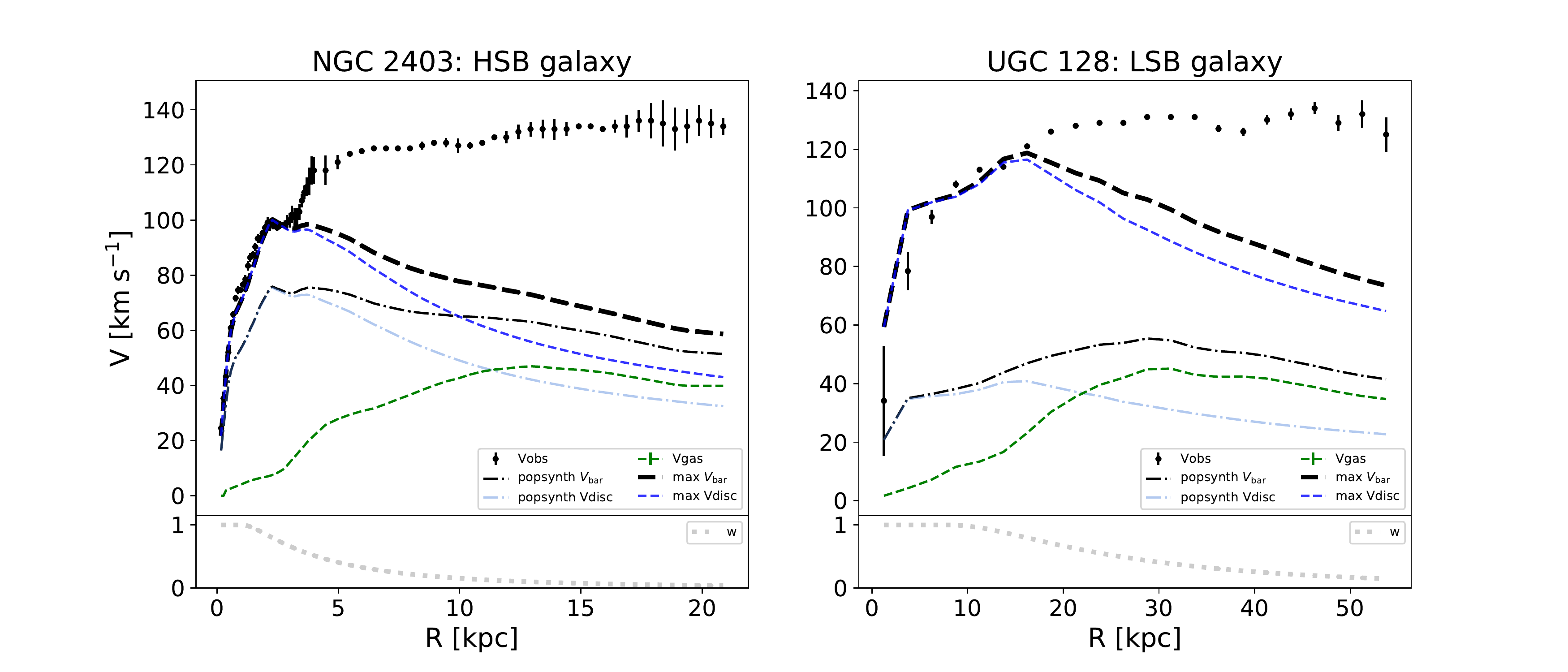}
	\caption{Examples of two pure disc galaxies. Left: The HSB galaxy NGC2403. Right: The LSB galaxy UGC\,128. Each panel shows the observed rotation curve (dots with error bars), the gas contribution (green dashed line), and the stellar disc contribution for two different choices of $\MtLdisc$: the maximum disc value (dashed blue line) and the pop-synth value of 0.5 $\rm{M}_{\odot}/\rm{L}_{\odot}$ from stellar population models (dot-dashed blue line). The corresponding total baryonic contributions are shown by the black dashed and dot-dashed lines, respectively. These cases illustrate a general trend: the disc of the HSB galaxy is maximized by a modest increase ($\lesssim$2) of the mass-to-light ratio above the nominal stellar population value, while that of the LSB galaxy requires a much larger boost in excess of a factor of 8. The small bottom panel shows the modified log-normal weighting function (dotted line) normalized to one. \label{fig:LSBHSBsidebyside}}
\end{figure*}

We also add a condition to under-weight over-maximal discs. When $\mathcal{F}(R)$ is negative (over-maximal discs), we increase its magnitude by a factor $f\cdot \mathcal{F}(R)$ where $f$ is a user-defined positive constant: $\mathcal{F}(R) \mapsto (1 + f)\cdot \mathcal{F}(R) < \mathcal{F}(R)$. This selects against overmaximal fits proportionally to their degree of maximality. Fits are insensitive to changes in $f$, provided $f \gtrsim \mathcal{O}(1)$. Larger values of $f$ suppresses fits where $\Vbary$ exceeds $\Vobs$ at any point, particularly when $\Vbary \approx \Vobs$ at large $R$. We adopt for $f$ the conservative value of unity.


\subsection{Best-Fit Values and Uncertainties}
\label{sub:best_fit_values_and_uncertainties}

The fit is obtained by minimizing the residual function with supermaximality condition, using either Nelder-Mead or least-squares algorithms. The minimizations are performed utilizing the LMFit library \citep{newville_stensitzki_allen_ingargiola_2014}, which expand upon SCIPY's optimization library \citep{SCIPY}. When a bulge is present, we perform a sequence of fits by iteratively alternating the two free parameter ($\MtLdisc$ and $\MtLbul$) to better quantify the full covariance. The resulting values are the \textit{maximal} mass-to-light of disc and bulge, denoted as $\maxMtLdisc$ and $\maxMtLbul$, respectively.

The uncertainties in the fit are given by the least-squares algorithm from the SCIPY's optimization package. However, when both disc and bulge are present, any minimization scheme struggles to obtain meaningful estimates of the uncertainties due to the degeneracy between $\MtLdisc$ and $\MtLbul$. To this end, the uncertainties are corroborated using a MCMC algorithm.

The MCMC fits use the LMFit functions \citep{newville_2014_11813}, which are wrappers for MCMC functions in the EMCEE library \citep{2013PASP..125..306F}. To train the MCMC, each fit has a burn-in period of 200 steps before taking 600 data points with a step-rate of 10 iterations between data point collections. Unavoidably, the MCMC is orders of magnitude slower than the Nelder-Mead and least-squares minimization, but allows a much more accurate uncertainty estimation. The median values of the MCMC often converge upon the Nelder-Mead and least-squares fit values.



\section{Results}
\label{sec:Results}

\begin{figure*}
	\centering
	\subfloat{%
		\includegraphics[width=0.45\textwidth]{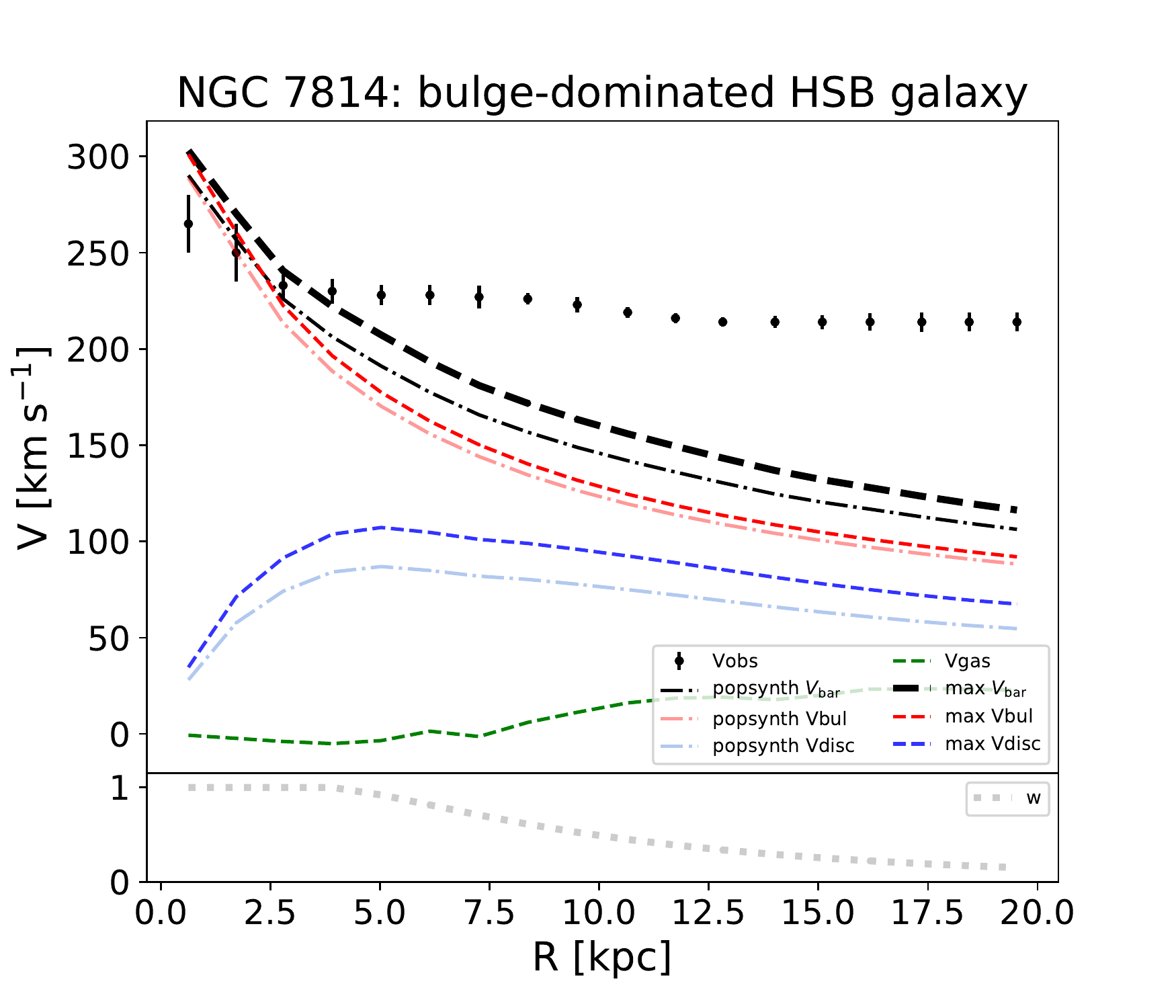}
	} \quad
 	\subfloat{%
		\includegraphics[width=0.45\textwidth]{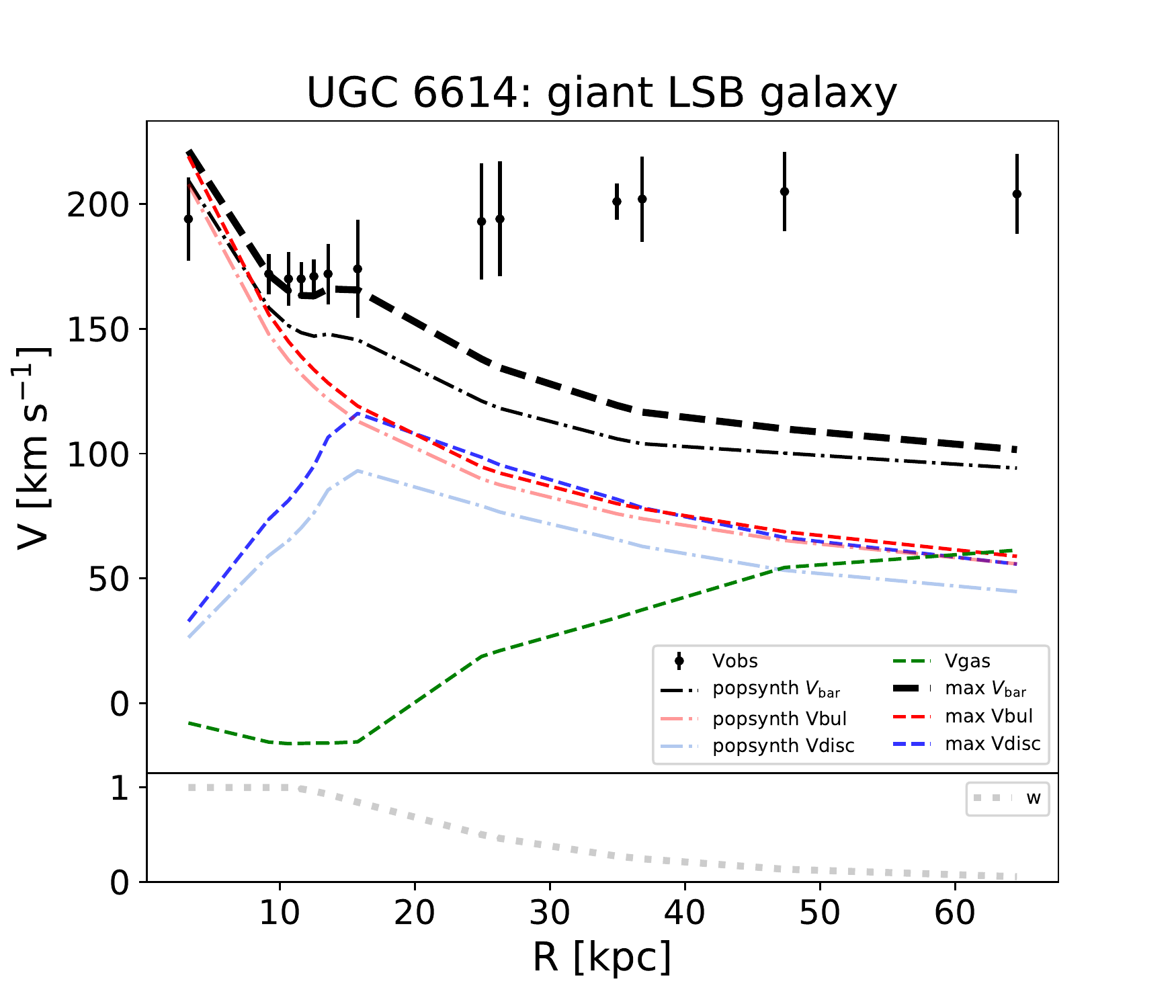}
	} \quad
	\caption{Examples of two bulge-dominated galaxies. Left: the HSB spiral NGC\,7814. Right: the LSB galaxy UGC\,6614. Lines and symbols are the same as in \autoref{fig:LSBHSBsidebyside}, but we also show the bulge contribution assuming the maximum value of $\MtLbul$ (red dashed line) and the pop-synth value of 0.7 $\rm{M}_{\odot}/\rm{L}_{\odot}$ from stellar population synthesis models (red dash-dotted line). \label{fig:bulgeexamples}}
\end{figure*}

The galaxies in the SPARC database span the broadest possible range of disc properties and display a large diversity in rotation curve profiles. Some are archetypal HSB spirals with rotation curves that rise fast and flatten at small radii. Others are LSB discs with slowly rising rotation curves that start to flatten only in the outer parts. Some galaxies have central bulges and their rotation curves display a very steep inner rise, a Keplerian-like intermediate decline, and a flat outer part. Some rotation curves are well sampled but others are not. Figures \ref{fig:LSBHSBsidebyside}-\ref{fig:dwarfexamples} present a selection of maximum disc fits to illustrate this diversity. Specifically, we show typical disc galaxies with HSB or LSB profiles (\autoref{sub:results_with_only_discs}), massive galaxies with small or large bulges (\autoref{sub:results_with_bulges}), and dwarf galaxies with well or sparsely sampled rotation curves (\autoref{sub:low_mass_dwarf_galaxies}).

\subsection{Pure Disc Galaxies}
\label{sub:results_with_only_discs}

In the following, we discuss galaxies with archetypal rotation curves such as the HSB galaxy NGC\,2403 and the LSB galaxy UGC\,128. This demonstrates the robustness of the maximization algorithm on typical galaxy discs.

\subsubsection*{NGC\,2403}
\label{ssub:ngc_2403}

\autoref{fig:LSBHSBsidebyside} (left) shows NGC\,2403 which is a classic example of an HSB spiral galaxy \citep{1986RSPTA.320..447V, 1997MNRAS.290..533D, 2002AJ....123.3124F}. NGC\,2403 has an archetypal flat rotation curve. It also displays a characteristic feature at $R\simeq3$ kpc, which is known to be reproduced by maximum disc fits \citep{1986RSPTA.320..447V, 2002AJ....123.3124F}. Our automatic algorithm provides the same result, so it can recover well classical visual fits with no manual intervention. The baryonic and observed rotation curves diverge past the plateau, pointing to a significant dark matter contribution beyond this radius. Since typical dark matter models do not contain central density holes, this maximal scaling is not physical but does provide a hard upper limit to the mass-to-light ratio of the disc. We find $\maxMtLdisc = 0.88 \pm 0.02 \, \rm{M}_{\odot}/\rm{L}_{\odot}$, which is acceptable in terms of stellar populations albeit slightly higher than the SPARC population synthesis (hereafter pop-synth) value of $0.5 \, \rm{M}_{\odot}/\rm{L}_{\odot}$. NGC\,2403 likely has a nearly maximal disc.



\subsubsection*{UGC\,128}
\label{ssub:ugc_128}

\autoref{fig:LSBHSBsidebyside} (right) shows UGC\,128, which is a LSB counterpart of NGC\,2403 \citep{Verheijen:1999fw}. The two galaxies have similar $V_{\rm flat}$ and total baryonic mass, but the effective radius of UGC\,128 is 4.5 times larger than that of NGC\,2403. Hence, the stellar and gas discs of UGC\,128 have significantly lower surface densities. This difference in surface densities is known to correspond to a difference in rotation curve shapes \citep{1997MNRAS.290..533D, 2013MNRAS.433L..30L}. The rotation curve of NGC\,2403 reaches $\Vflat\simeq130$ km\,s$^{-1}$ within the first $\sim$5 kpc, while the one of UGC\,128 has barely reached the same value of $\Vflat$ within 40 kpc. When $\MtLdisc$ is maximized, $\Vdisc$ can reproduce the rotation curve until $\sim$16 kpc, which is comparable NGC\,2403 when scaled by the disc scale length $\Rd$ ($16/4.5\simeq3.5$ kpc). There is, however, a major difference. UGC\,128 has $\maxMtLdisc=4.07 \pm 0.17$, significantly larger than NGC\,2403. It is well-known that LSB galaxies have larger $\maxMtLdisc$ than HSB ones \citep{1997MNRAS.290..533D}. Hence, our automatic fitting algorithm is able to reproduce another classic result that was derived by manually adjusting the stellar mass-to-light ratios. The general trend between $\maxMtLdisc$ and surface brightness is discussed in detail in \autoref{sec:Discussion}.

The maximal disc fits to NGC\,2403 and UGC\,128 are fairly typical for disc galaxies in the SPARC database. In most such galaxies, maximum disc fits give  $\Vdisc > \Vgas$ at small radii, while $\Vgas$ can become almost dominant at larger radii. Moreover, the maximal baryonic rotation curve tracks the observed rotation curve over a range of radii, suggesting a coupling between baryons and dynamics.


\begin{figure*}
	\flushright
 	\subfloat{%
		\includegraphics[width=1.0\textwidth]{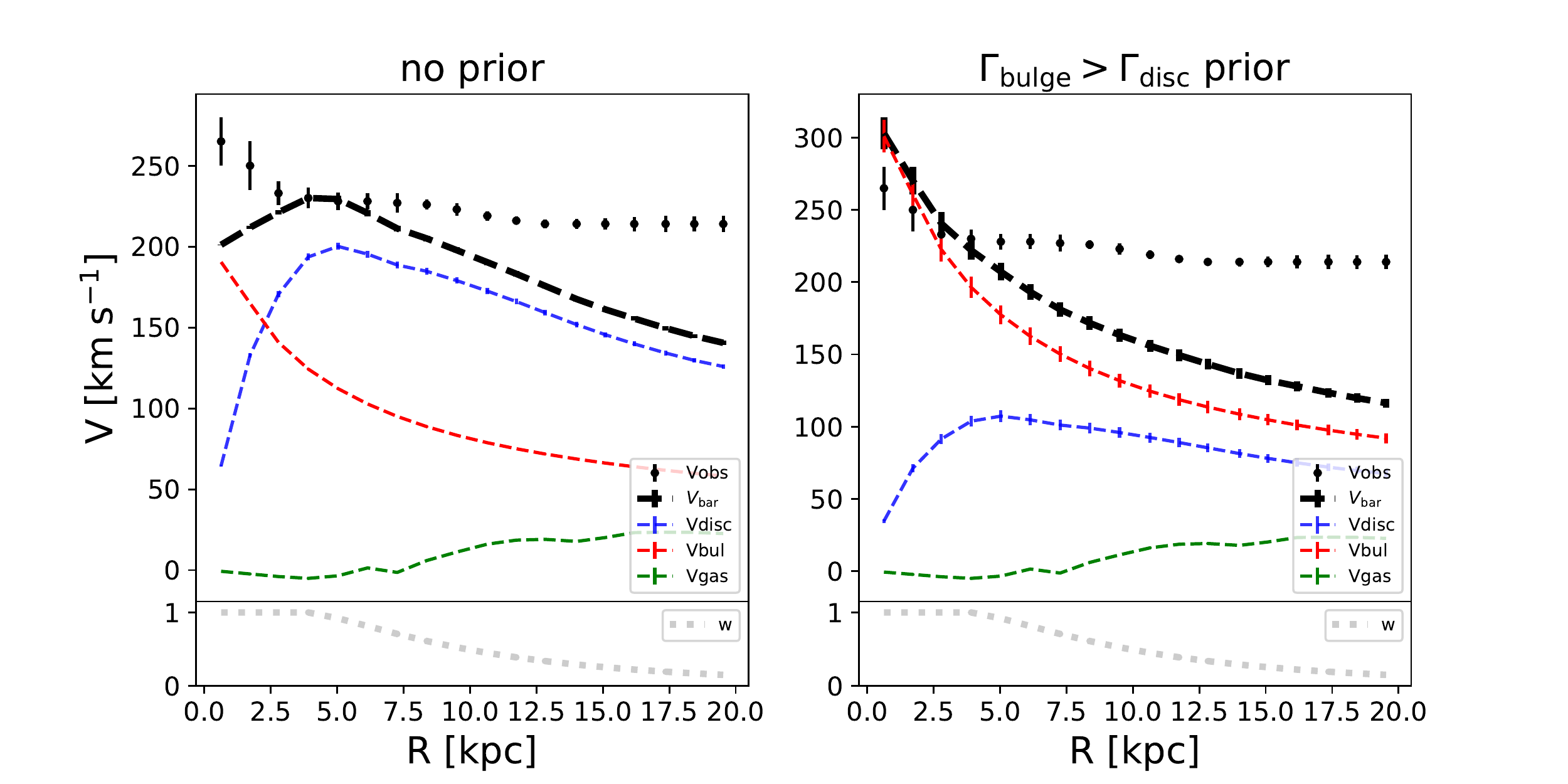}
	} \\
	\subfloat{%
		\includegraphics[width=0.42\textwidth]{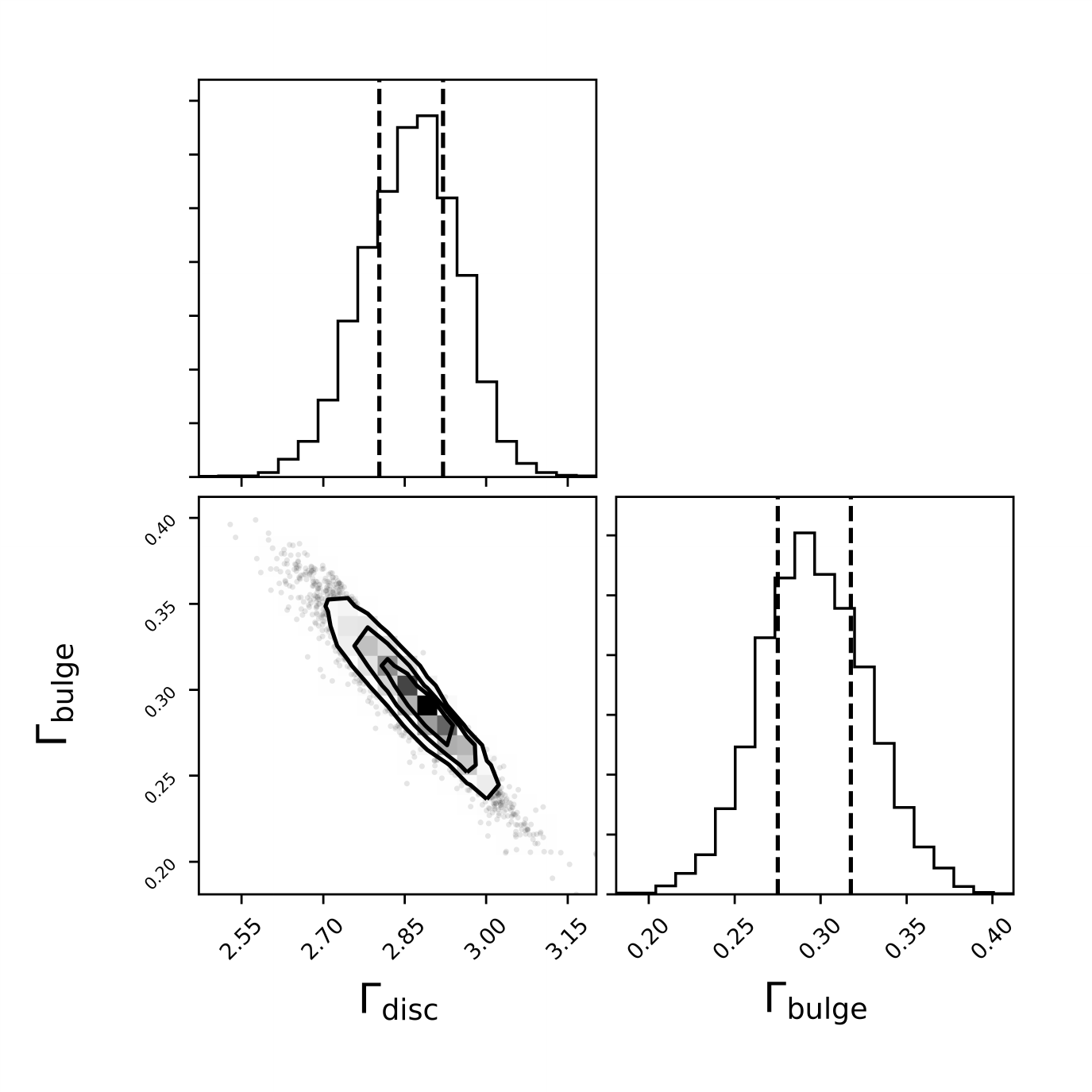}
	}
	\subfloat{%
		\includegraphics[width=0.42\textwidth]{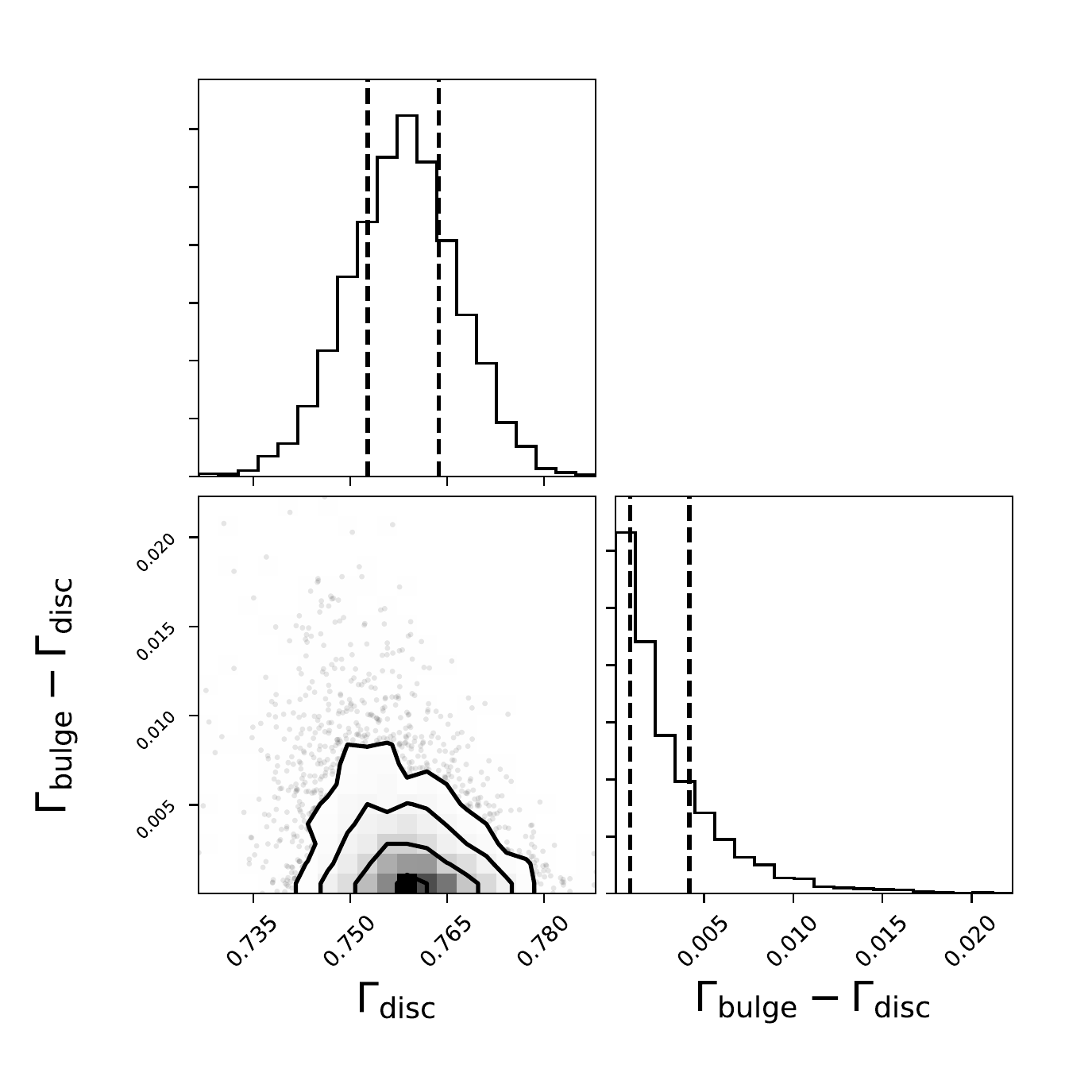}
	} \hspace{35pt}
	\caption{Examples of fits to the galaxy NGC\,7814. Left: the maximal fit with no additional priors besides the weighting function and $0<\maxMtLdisc,\maxMtLbul<20$. $\Vbary$ is not maximal in the bulge-dominated region of $\Vobs$. Right: the maximal fit with the additional prior $\maxMtLbul>\maxMtLdisc$. The lower panels show the posterior distribution of fitting parameters from the MCMC fits. $\MtLbul$ and $\MtLdisc$ are degenerated but a clear minimum is found in the right panel. We use lax priors: the mass-to-light ratios must be positive and smaller than 20, imposing in the right panel the additional constraint $\maxMtLbul > \maxMtLdisc$.\label{fig:ngc_7814}}
\end{figure*}

\subsection{Galaxies with Bulges}
\label{sub:results_with_bulges}
Galaxies with bulges pose a challenge to devising an objective maximum disc algorithm. There is considerable degeneracy between bulge and disc: maximizing one degrades the other. Consequently, different choices for the priors can yield qualitatively different results. In the following, we present two examples of galaxies with bulges: the typical HSB galaxy NGC\,7814 and the more rare giant LSB galaxy UGC\,6614. The former is used to illustrate the degeneracy between $\MtLbul$ and $\MtLdisc$ as well as how this degeneracy can be broken. Bulge-dominated galaxies are a minority in the SPARC database and occur mostly in the HSB regime.

\subsubsection*{NGC\,7814} 
\label{ssub:ngc_7814}

\autoref{fig:bulgeexamples} (left) shows the classic bulge-dominated HSB spiral NGC\,7814. Our automated algorithm gives $\MtLbul = 0.76\pm 0.05$ and $\MtLdisc = 0.76\pm 0.06$, which are acceptable in terms of stellar populations. The bulge component nicely reproduces the inner declining portion of the rotation curve and dominates over the disc out to the last measured point.

In \autoref{fig:ngc_7814} (top panels), we show the key role of the $\maxMtLbul > \maxMtLdisc$ prior to achieve sensible fits. Without this prior (left), the algorithm reproduces $\Vobs$ at $3 \lesssim R \lesssim 6$ kpc but fails to reproduce the inner points at $R\lesssim3$ kpc. This happens because the declining portion of the rotation curve is not well sampled, so the weighting function $w(R)$ prioritizes the disc over the bulge to achieve baryonic maximality out to the largest possible radii. The bulge has no room left to reproduce the inner declining portion of the rotation curve: increasing $V_{\rm bul}$ would overshoot $V_{\rm obs}$ at intermediate radii.

In \autoref{fig:ngc_7814} (bottom panels), we show the posterior distributions of the fitting parameters. Without the $\maxMtLbul > \maxMtLdisc$ prior (left), the two parameters are degenerated. A minimum is present but it is clear that this cannot be physical: the inner declining portion of the rotation curve must be driven by the bulge, given the similar shape of $V_{\rm obs}$ and $V_{\rm bul}$ at $R\lesssim 3$ kpc. The $\maxMtLbul > \maxMtLdisc$ prior (right) can break the parameter parameter degeneracy and give a truly maximal bulge.

We note that the bulge-disc degeneracy happens in most galaxies with bulges despite the careful bulge-disc decompositions of the luminosity profiles. The degeneracy lies in the conversion from light to mass, not in the underlying luminosity decomposition.


\subsubsection*{UGC\,6614} 
\label{ssub:ugc_6614}
UGC\,6614 belongs to the class of giant LSB galaxies \citep{1995AJ....109..558S} (\autoref{fig:bulgeexamples}, left). Giant LSB galaxies have very extended LSB discs and central light concentrations, which can be due to either a single large bulge \citep{MSB95} or a more complicated HSB structure, like a small bulge plus a compact disc \citep{2007AJ....133.1085B, 2010A&A...516A..11L}. In the case of UGC\,6614, bulge and disc are not strongly degenerated because the rotation curve shows an inner decline driven by the bulge and a subsequent rising portion driven by the disc. These specific trends in the rotation curve shape prevents the degeneracy between $\MtLdisc$ and $\MtLbul$.

When disc and bulge are maximized, $\Vbary$ can explain the observed rotation curve out to $\sim$20 kpc. The bulge dominates out to $\sim$15 kpc. The maximal mass-to-light ratios are $\maxMtLdisc = \maxMtLbul = 0.77 \pm 0.05$. The former value is near-identical to stellar population synthesis models. The latter is a modest increase from the pop-synth value of $0.5 \, \rm{M}_{\odot}/\rm{L}_{\odot}$. This is in line with the picture where giant LSB have a double dynamical structure: an inner baryon-dominated HSB region and an outer DM-dominated portion \citep{2010A&A...516A..11L}.



\subsection{Low-Mass Dwarf Galaxies}
\label{sub:low_mass_dwarf_galaxies}

\begin{figure*}
	\centering
	\makebox[\textwidth][c]{\includegraphics[width=1.2\linewidth]{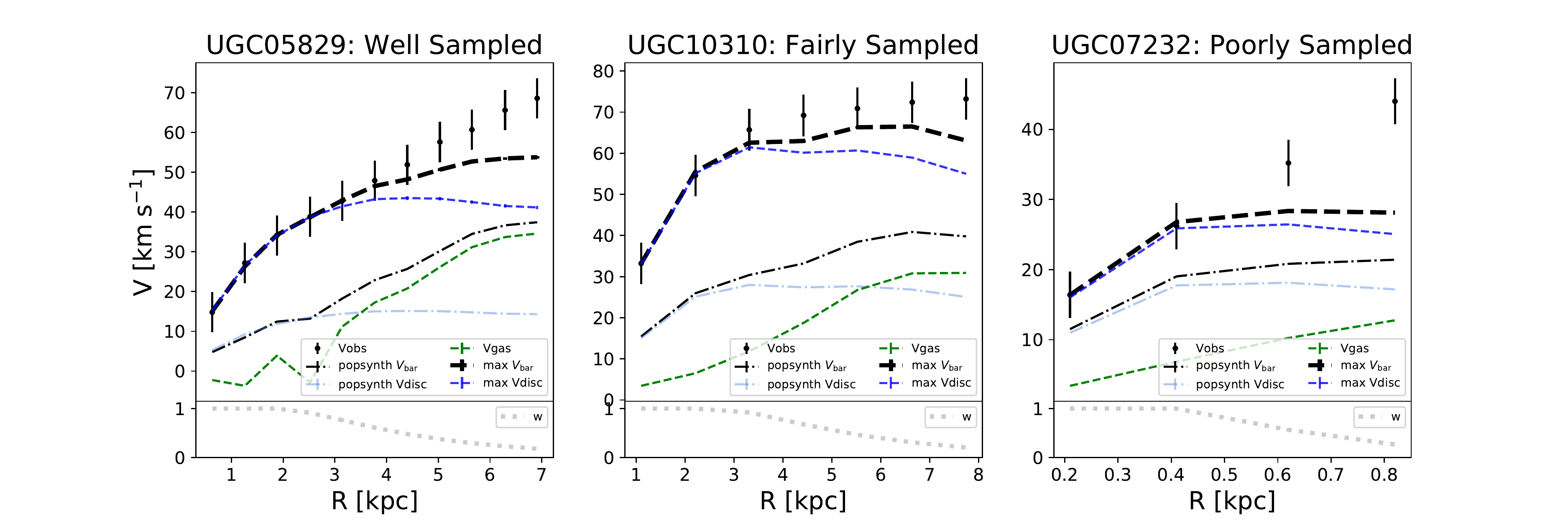}}
	\caption{Examples of dwarf galaxies with differently sampled rotation curves: UGC\,5829 (left, well-sampled), UGC\,10310 (middle, fairly sampled), and UGC\,7232 (right, poorly sampled). Lines and symbols are the same as in \autoref{fig:LSBHSBsidebyside}.} \label{fig:dwarfexamples}
\end{figure*}

Low-mass, dwarf galaxies are generally bulgeless and the maximization involves a single parameter, as in pure disc galaxies (\autoref{sub:results_with_only_discs}). Dwarf galaxies, however, can be heavily gas dominated having $M_{\rm gas} > M_{\rm star}$ for pop-synth stellar mass-to-light ratios. In some cases, this implies that the maximum disc fit is rather insensitive to the exact value of $\Upsilon_{\rm star}$, especially when $V_{\rm gas}$ can already explain $V_{\rm obs}$ at the innermost radii. In other cases, the maximum disc fits may wash out trends and features that are observed in both $\Vgas$ and $\Vobs$, but are not present in $\Vdisc$. As a result, maximizing the stellar disc of dwarfs produces varied results: for some objects $\Vobs$ may be fully explained by the baryons, while for others $\Vbary$ scarcely changes from the sub-maximal stellar population value.

Given their small size on the sky, dwarf galaxies can have sparsely sampled rotation curves. In the following, we describe three dwarf galaxies spanning a range of sampling quality: UGC\,5829 (well sampled), UGC\,10310 (fairly sampled), and UGC\,7232 (poorly sampled).

\subsubsection*{UGC\,5829} 
\label{ssub:ugc_5829}

\autoref{fig:dwarfexamples} (left) shows a dwarf galaxy with a well sampled rotation curve: UGC\,5829. The observed rotation curve continues to rise out to the last measured point. This is well after the stellar contribution has peaked at $\sim$4 kpc. Interestingly, the gas contribution also continues to rise to the last measured point. When the maximal scaling is applied, $\Vbary$ matches $\Vobs$ until $\sim$4.5 kpc. At large radii, dark matter is absolutely necessary, albeit one may also scale up the gas component to fully reproduce the rotation curve \citep[see][]{2012MNRAS.425.2299S}.
Maximizing the stellar disc requires $\MtLdisc = 4.14 \pm 0.08$: this is an eight-fold increase with respect to the pop-synth value of $0.5 \, \rm{M}_{\odot}/\rm{L}_{\odot}$, which seems unphysical. This suggests that UGC\,5829 does not have a maximal disc.


\subsubsection*{UGC\,10310}
\label{ssub:ugc_10310}

\autoref{fig:dwarfexamples} (middle) shows a dwarf galaxy with a fairly sampled rotation curve: UGC\,10310. The rotation curve reaches a flat part at $\sim$5 kpc. When the stellar disc is maximized, it can almost entirely explains $\Vobs$. This is unusual for dwarf galaxies. This is unrelated to the sampling of the rotation curve, but it is due to the fact that $\Vdisc$ is relatively flat and featureless at large radii, so it preserves the salient features in $\Vgas$. The maximum $\maxMtLdisc$ is $2.41 \pm 0.06$, which again exceeds the expectations from standard stellar population models.


\subsubsection*{UGC\,7232}
\label{ssub:maximizing_dwarfs}

\autoref{fig:dwarfexamples} (middle) shows a dwarf galaxy with a fairly sampled rotation curve: UGC\,7232.
This object has only four velocity points. While this does not impact fit quality per se, it makes difficult the comparisons with better sampled galaxies. Poorly sampled galaxies like UGC\,7232 often appear as outliers in statistical analyses of maximal disc fits (\autoref{sub:the_maximal_mass_to_light_ratio}). This does not necessarily mean that they are atypical (though they may be) but it likely indicates that the underlying data is incomplete.

\section{Discussion} 
\label{sec:Discussion}

\subsection{The Maximal Mass-to-Light Ratio}
\label{sub:the_maximal_mass_to_light_ratio}

\autoref{table:disc_maximality} provides the fit results for the example galaxies in Figures \ref{fig:LSBHSBsidebyside} to \ref{fig:dwarfexamples}, ordered by increasing stellar surface density. Fit results for all 175 SPARC galaxies are provided in machine readable form at astroweb.cwru.edu/SPARC. This galaxy selection illustrates the general difference between HSB and LSB galaxies. For example, the HSB galaxies NGC\,2403 and NGC\,7814 have $\maxMtLdisc$ equal to 0.88 and 0.76, respectively. These values are close to expectations of stellar population synthesis models ($\sim 0.5 \, \rm{M}_{\odot}/\rm{L}_{\odot}$), albeit slightly larger. This is expected because our maximum disc definition allows no room for dark matter in the center, so it is conceivable that the actual stellar mass-to-light ratios will be somewhat smaller than $\maxMtLdisc$ and $\maxMtLbul$. Figures \ref{fig:LSBHSBsidebyside} and \ref{fig:bulgeexamples} show that the difference between $0.5$ and $0.8 \ \rm{M}_{\odot}/\rm{L}_{\odot}$ is generally small for HSB galaxies.

On the other hand, LSB galaxies like UGC\,5829 and UGC\,128 have $\maxMtLdisc = 4.14$ and $\maxMtLdisc = 4.07$, respectively, which are $\sim$8 times larger than expected from stellar population synthesis models. It is evident that LSB galaxies cannot be maximal. To produce these high stellar mass-to-light ratios, one would need either extremely high star formation activity in the early Universe or an extremely bottom-heavy IMF. Both options contrast our understanding of LSB galaxies as slowly evolving systems.

\begin{table}
	\centering
	\ra{1.3} \setlength\tabcolsep{2.5 pt} \fontsize{9}{9}\selectfont
	\begin{tabular}{ l l l c c c c }
		\hline
		ID & $\log_{10}\SBbar$ & T & $\maxMtLdisc$ & $\maxMtLbul$ & $\VbVp$ \\
		   & $\rm{M}_\odot \; \rm{pc}^{-2}$    &   & ${\rm{M}_\odot} {\rm{L}_\odot}^{-1}$ & ${\rm{M}_\odot} {\rm{L}_\odot}^{-1}$ & \\
		\hline

		UGC\,5829  & $1.56 \spacepm 0.02  $ & 10  & $4.14\spacepm0.08$ &  -   & $0.78\spacepm0.06$ \\
		UGC\,128   & $1.65 \spacepm 0.03  $ & 8   & $4.07\spacepm0.18$ &  -   & $0.91\spacepm0.02$ \\
		UGC\,10310 & $1.85 \spacepm 0.01  $ & 9   & $2.41\spacepm0.06$ &  -   & $0.92\spacepm0.06$ \\
		UGC\,7232  & $2.37 \spacepm 0.01   $ & 10  & $1.06\spacepm0.13$ &  -   & $0.81\spacepm0.09$ \\
		NGC\,6946  & $5.91 \spacepm 0.01   $ & 6   & $0.59\spacepm0.01$ & $0.59\spacepm0.01$ & $1.18\spacepm0.06$ \\
		UGC\,2885  & $4.97 \spacepm 0.02  $ & 5   & $0.89\spacepm0.04$ & $0.89\spacepm0.04$ & $0.87\spacepm0.03$ \\
		NGC\,2403  & $3.10 \spacepm 0.001   $ & 6   & $0.88\spacepm0.02$ &  -   & $1.02\spacepm0.02$ \\
		NGC\,891   & $3.62 \spacepm 0.04  $ & 3   & $0.43\spacepm0.04$ & $0.43\spacepm0.04$ & $1.11\spacepm0.05$ \\
		UGC\,6614  & $3.84 \spacepm 0.03  $ & 1   & $0.78\spacepm0.08$ & $0.78\spacepm0.08$ & $1.14\spacepm 0.10$ \\
		NGC\,7814  & $5.03 \spacepm 0.03$ & 2   & $0.76\spacepm0.06$ & $0.76\spacepm0.06$ & $1.14\spacepm0.08$ \\
		\hline

	\end{tabular}
 	\caption{Maximum disc parameters for the example galaxies in \autoref{fig:LSBHSBsidebyside}, \autoref{fig:bulgeexamples}, \autoref{fig:ngc_7814}, and \autoref{fig:dwarfexamples}. The full table with 153 galaxies is available in electronic form at astroweb.cwru.edu/SPARC.\label{table:disc_maximality}}
\end{table}

\begin{figure*}
	\includegraphics[width=.8\linewidth]{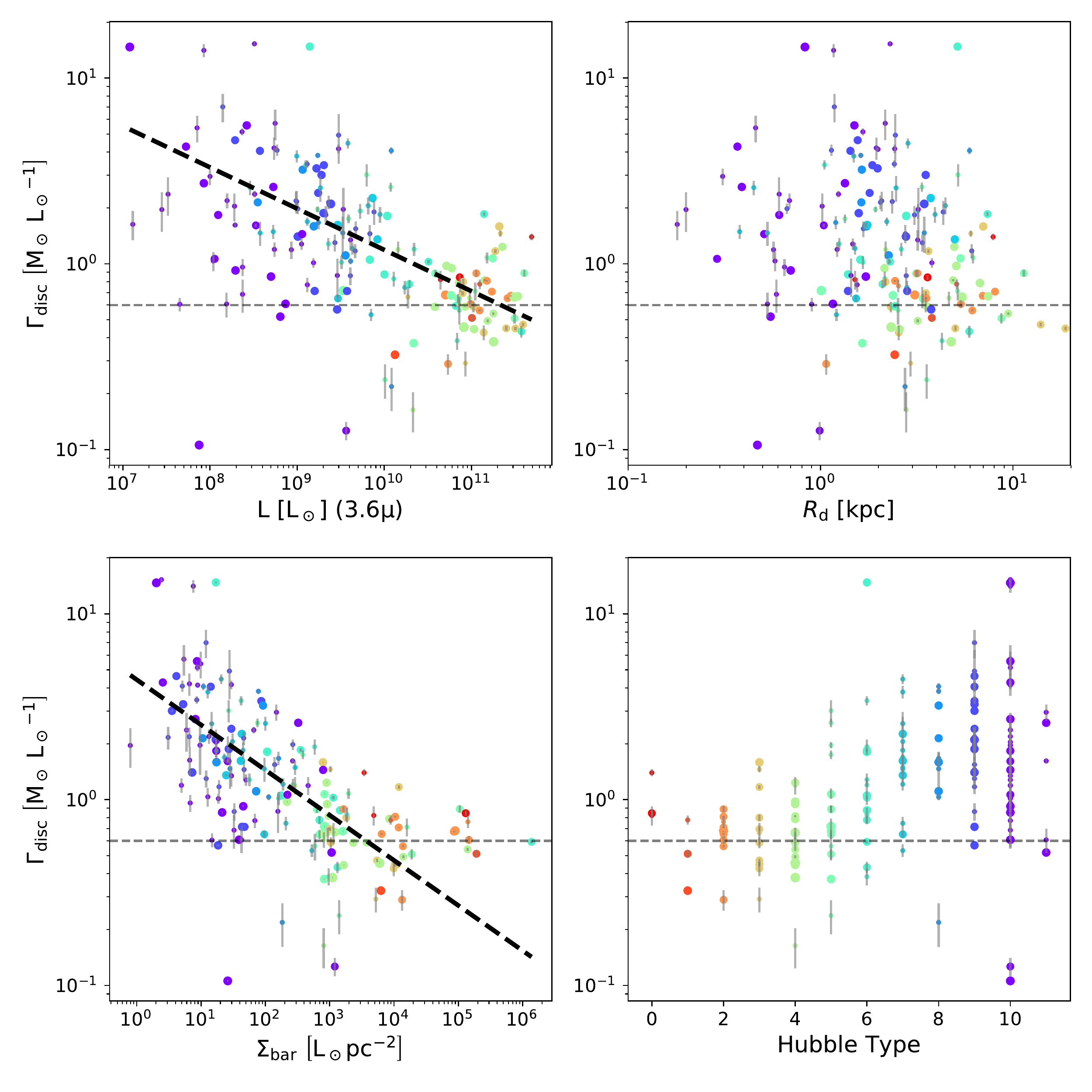}
	\caption{The maximum-disc mass-to-light ratio ($\maxMtLdisc$) versus luminosity at 3.6 $\mu$m (top left), disc scale-length (top right), surface brightness (bottom left), and Hubble type. The surface brightness has units of solar masses per square parsec and is constructed with the pop-synth model. $\Rd$ is in kpc. Luminosity is given in $10^{9} \; \rm{L}_\odot$ as measured by the Spitzer 3.6 micron band. Data points are scaled inversely to error size and colour coded by galaxy Hubble type. Least-squares fit between $\maxMtLdisc$ and surface brightness shows correlation, and large scatter, with $\log_{10}(\maxMtLdisc) = 0.57 - 0.24 \log_{10}(\SBbar \ [{\rm{L}_\odot}{\rm{pc}^{-2}}])$. There is a similar, albeit weaker, correlation with luminosity. The dashed gray line shows $\maxMtLdisc=0.6$, the midpoint and rough separating value between the population synthesis values for non-maximal discs and near-maximal bulges.\label{fig:m2l_correlations}}
\end{figure*}

\autoref{fig:m2l_correlations} shows the maximum $\maxMtLdisc$ versus several galaxy properties: total luminosity at 3.6 $\mu$m (top left), disc scale length (top right), effective surface brightness (bottom left), and Hubble type. The values of $\maxMtLdisc$ spans a broad range of values, from $\sim$0.1 to $\sim$10. The median $\maxMtLdisc$ is $\sim$1.2 with inner quartiles 0.7 and 2.0. These values are higher than pop-synth values from stellar population models, but are clearly inflated by low-luminosity and LSB galaxies. Indeed, $\maxMtLdisc$ correlates with both luminosity and effective surface brightness. A linear fit gives:

\begin{equation}\label{eqn:maxMtLdisc_to_L}
\log_{10}(\maxMtLdisc) = 0.30 - 0.22 \log_{10}(L_{3.6})
\end{equation}
with a rms scatter of 0.3 dex, and
\begin{align}
\log_{10}(\maxMtLdisc) &= 0.65 - 0.24 \log_{10}(\SBbar \ [{\rm{M}_\odot}{\rm{pc}^{-2}}]) \label{eqn:maxMtLdisc_to_SBbar}\\
\log_{10}(\maxMtLdisc) &= 0.57 - 0.24 \log_{10}(\mathcal{E}_{\rm{bar}} \ [{\rm{L}_\odot}{\rm{pc}^{-2}}]) \nonumber
\end{align}
likewise with a rms scatter of 0.3 dex.

These correlations can be understood in terms of the systematic change in dark matter domination with luminosity and surface brightness. Considering the value of $\MtLdisc$ from stellar population models as a baseline starting point, LSB galaxies scale up their discs more than HSB galaxies do. If we assume that the actual stellar mass-to-light ratios of LSB and HSB galaxies are similar, LSB galaxies must be more dark matter dominated than HSB galaxies. In other words, LSB galaxies necessitate larger $\maxMtLdisc$ because they are correspondingly more dark matter dominated.

\subsection{Degree of Maximality} 
\label{sub:degree_of_maximality}

Historically, maximum-disc mass models have been build by scaling up the stellar contribution ``by hand'', such that the peak of the baryonic rotation curve is comparable to the peak of the observed rotation curve. For a pure exponential disc, the peak of the baryonic rotation curve occurs at $\sim$2.2 disc scale lengths ($\Rd$), so it is customary to measure the degree of maximality using the ratio $\VbVp$ at 2.2 $\Rd$. \citet{1997ApJ...483..103S} analysed maximum disc decompositions from previous authors and found that $\VbVp = 0.85 \pm 0.10$ at 2.2 $\Rd$. This is often used as a definition for the maximum disc, despite a truly maximum disc should have $\VbVp = 1$. The reasons are twofold: (i) visual maximum disc decompositions were performed leaving some ``room'' for dark matter near the galaxy center in order to avoid hollow halos, (ii) low-mass galaxies show a different behaviour than high-mass ones, which leads to $\VbVp <1$.

In many galaxies, however, $\Vb$ does not peak at 2.2 $\Rd$ because either the stellar distribution deviates from an exponential profile (such as in bulge-dominated galaxies) or the gas contribution fully dominates the baryonic budget (such as in some dwarf galaxies). Thus, it is preferable to measure $\VbVp$ at $\Rbary$, the radius where the total baryonic rotation curve peaks \citep[see][]{2016AJ....152..157L}.

\begin{figure*}
    \includegraphics[width=.8\linewidth]{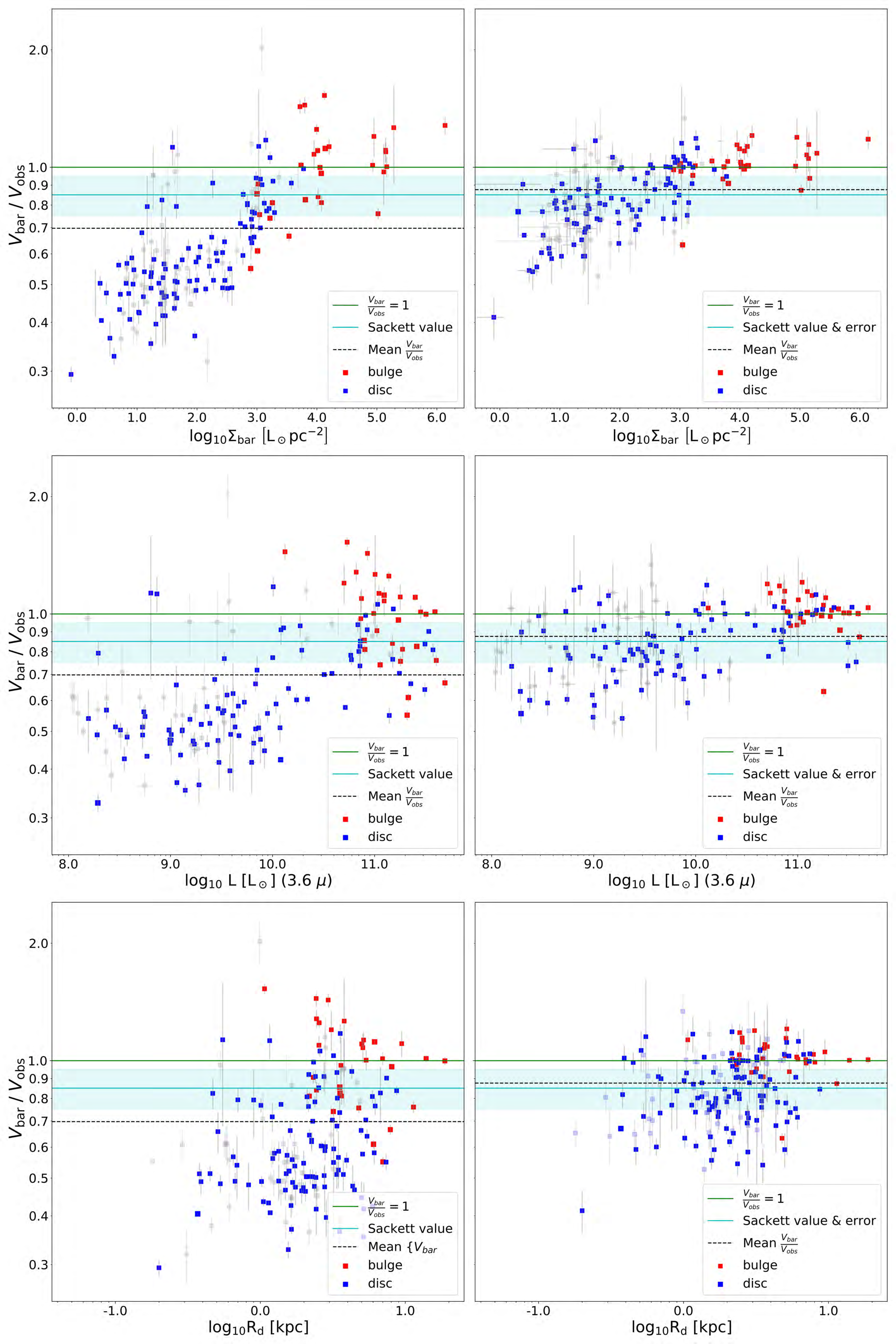}
    \caption{$\VbVp$ vs $\SBbar$, luminosity, and $\Rd$ for all galaxies in SPARC. Left: disc models assuming a constant $\MtLdisc = 0.5$, $\MtLbul = 0.7$. Right: maximized ($\maxMtLdisc$, $\maxMtLbul$) disc models. Shaded green shows $\VbVp$ saturation area for maximized discs. Shaded cyan shows the value obtained in \citet{1997ApJ...483..103S}. This is compared with the median value, show in black. Galaxies with bulges are in red, while disc-only are in blue. Galaxies in gray have fewer than 8 data points in $\Vobs$, so the peaks of $\Vbary$ and $\Vobs$, and $\Rbary$ are not well defined. The $\SBbar$ correlation is the tightest shown for the pop-synth model. Post maximizations, the low surface brightness galaxies remain correlated with $\VbVp$. Most LSB galaxies which are near maximal have rotation curves not sampled to the peak, meaning $\VbVp$ measures in the inner regions, where the disc is maximal.  \label{fig:VbVp_4panel_ln_fit_prefit_comparison}}
\end{figure*}

\autoref{fig:VbVp_4panel_ln_fit_prefit_comparison} shows $\VbVp$ at $\Rbary$ versus several galaxy properties: surface brightness (top), luminosity at 3.6 $\mu$m (middle), and $\Rd$ (bottom). The left and right columns correspond to values of $\Upsilon_{\star}$ from stellar populations synthesis models and maximum-disc fits, respectively. After applying the maximum disc procedure, the $\VbVp$ values of several low-mass and LSB galaxies increase with respect to the stellar population expectation, albeit they do not necessarily reach $\VbVp = 1$. This occurs because $V_{\rm bar}$ is often dominated by the gas component and peaks at large radii, where some discrepancy between $\Vb$ and $V_{\rm obs}$ is unavoidable for any value of $\Upsilon_{\star}$ (see \autoref{fig:dwarfexamples}, left). On the other hand, the $\VbVp$ values of some high-mass galaxies (especially the ones with bulges) decrease from unphysical values ($\VbVp > 1$) to truly maximal ones ($\VbVp \simeq 1$). This indicates that our condition against over-maximality works properly (see \autoref{sub:residual_function}).

A clear lesson from \autoref{fig:VbVp_4panel_ln_fit_prefit_comparison} is that the maximum-disc concept is not well described by the mean value of $\VbVp$. This quantity correlates with surface brightness and saturates at a value of $\sim$1 beyond $\sim 5 \times 10^3$ $L_{\odot}$ pc$^{-2}$, when bulges start to dominate the inner dynamics. Nevertheless, the mean value of $\VbVp$ is $0.88 \pm 0.06$, which is consistent with the classic value from \citet{1997ApJ...483..103S}. In our opinion, the value of $\VbVp$ should not be used to define maximum discs, especially if we aim to have a common definition for HSB and LSB galaxies. Our automatic algorithm overcomes this issue, providing an objective quantification (see \autoref{eqn:maxMtLdisc_to_L} and \autoref{eqn:maxMtLdisc_to_SBbar}) of ``maximum discs'' for all galaxy types.

\section{Conclusions} 
\label{sec:Conclusions}

We developed an algorithm to perform automatic maximum-disc fits to galaxy rotation curves. We applied our algorithm to 153 galaxies from the SPARC database and derived the maximum stellar mass-to-light ratios of bulge ($\maxMtLbul$) and disc ($\maxMtLdisc$). We showed that our procedure recovers, with no manual intervention, classic results from visual maximum-disc fits. Our main results can be summarized as follows:
\begin{itemize}
\item Maximum-disc fits to HSB galaxies return values of $\maxMtLbul$ and $\maxMtLdisc$ comparable to the expectations of stellar population synthesis models (albeit slightly higher on average), suggesting that HSB galaxies are close to be maximal.
\item Maximum-disc fits to LSB galaxies imply unphysically large stellar mass-to-light ratios, suggesting that LSB galaxies are dominated by dark matter in the inner parts.
\item $\maxMtLdisc$ correlates with Hubble type, galaxy luminosity, and surface density. These correlations can be explained by the systematic increase of dark matter domination in the inner parts of late-type, low-luminosity, LSB galaxies.
\item The mean and median value of $\VbVp$ is $0.88 \pm 0.06$, comparable to classic result of \citet{1997ApJ...483..103S}. However, $\VbVp$ is positively correlated with galaxy surface brightness, so its mean value has limited meaning and should not be used to define the ``maximum disc'' concept.
\end{itemize}
Our automated procedure can be easily applied to large galaxy samples. In the near future, thousands of rotation curves will be available from wide-field HI surveys with next-generation radio telescopes like ASKAP, APERTIF, and ultimately SKA. Thus, it is crucial to have well-tested automatic tools to analyse these massive datasets.


\section*{Acknowledgments}

This work is partially supported by CWRU SOURCE.




\bibliographystyle{mnras}
\bibliography{MaxDiskBib}





\bsp	
\label{lastpage}
\end{document}